\documentclass[twocolumn,showpacs,preprintnumbers,amsmath,amssymb]{revtex4}

\usepackage{graphicx}
\usepackage{dcolumn}
\usepackage{bm}

\begin{document}

\preprint{APS/123-QED}

\title{Polarization squeezing of intense pulses \\ with a fiber Sagnac interferometer}

\author{Joel Heersink}

\email{joel.heersink@optik.uni-erlangen.de}

\author{Tobias Gaber}
\author{Stefan Lorenz}
\author{Oliver Gl\"ockl}
\author{Natalia Korolkova}
\author{Gerd Leuchs}

\affiliation{Zentrum f\"ur Moderne Optik, Physikalisches Institut,\\
 Universit\"at Erlangen-N\"urnberg, Staudtstra\ss e 7/B2, 91058 Erlangen, Germany.}


\date{\today}

\begin{abstract}
We report on the generation of polarization squeezing of intense, short light pulses using an asymmetric fiber Sagnac interferometer. The Kerr nonlinearity of the fiber is exploited to produce independent amplitude squeezed pulses. The polarization squeezing properties of spatially overlapped amplitude squeezed and coherent states are discussed. The experimental results for a single amplitude squeezed beam are compared to the case of two phase-matched, spatially overlapped amplitude squeezed pulses. For the latter, noise variances of -3.4~dB below shot noise in the $\hat{S}_0$ and the $\hat{S}_1$ and of -2.8~dB in the $\hat{S}_2$ Stokes parameters were observed, which is comparable to the input squeezing magnitude. Polarization squeezing, that is squeezing relative to a corresponding polarization minimum uncertainty state, was generated in $\hat{S}_1$.
\end{abstract}

\maketitle

\section{Introduction}

The nonclassical polarization states of light have recently been the subject of a number of theoretical and experimental papers in the framework of quantum information \cite{HalSor01,KorLeu01a,BowSch02,BowTre02}. Polarization squeezing was first discussed by Chirkin \cite{ChiOrl93}. This theoretical proposal used the nonlinear cross Kerr ($\chi^3$) effect with special requirements on the nonlinear coefficients to achieve noise reduction in polarization variables, i.e. the Stokes operators. The first experimental realization of polarization squeezing \cite{HalSor01} used an elegantly simple scheme where a squeezed vacuum was spatially overlapped with an orthogonally polarized strong coherent beam on a 50:50 beamsplitter. This experiment sought to map the polarization state of light, an information carrier, onto the spin state of a macroscopic atomic ensemble, which could be used as a quantum memory or as a quantum information processor \cite{HalSor01}. Prior to this experiment in the late 80's, the combination of squeezed vacuum and coherent beam of orthogonal polarization was used to enhance the sensitivity of a polarization interferometer for high precision phase measurements \cite{GraSlu87}. However, at that time the nonclassical polarization properties of the light field used were not recognized and results were attributed to the sub-shot-noise vacuum input, i.e. single-mode quadrature squeezing.


The interest in nonclassical polarization states has grown recently motivated by: 1) The demonstration of the mapping of quantum state of light onto atoms \cite{HalSor01}; 2) The recognition that the Stokes operators as conjugate variables which can all be measured in direct detection, rendering local oscillator methods unncessary \cite{KorLeu01a,KarMas93}; 3) The presentation of the concept of and the criteria for the experimental verification of continuous variable polarization entanglement \cite{KorLeu01a}. A straightforward experimental scheme for the generation of polarization squeezing and entanglement of intense beams was suggested in \cite{KorLeu01a}. Polarization squeezed light is generated by phase-locking two orthogonally polarized intense amplitude squeezed beams. Continuous variable polarization entanglement emerges from the linear interference of two polarization squeezed beams. In both schemes, the required nonclassical resources can be reduced to as few as one amplitude squeezed beam by substituting coherent or vacuum beams for the others. Depending on the particular combination, this degrades the degree of squeezing or entanglement. This deterioration may be weaker when using low intensity squeezed light \cite{BowSch02,BowTre02} rather than intense squeezed beams (see results in this paper). Using such low intensity beams, polarization squeezing of continuous wave beams was recently generated exploiting an optical parametric amplifier \cite{BowSch02} and was extensively experimentally characterized \cite{SchBow03}. In this paper we present results of a polarization squeezing experiment with pulsed light and devote special attention to the definition of polarization squeezing.

The paper is organized as follows: Sec. 2 briefly introduces the relevant polarization variables, the Stokes operators. The notion of polarization squeezing is also discussed and its specific properties are compared to conventional quadrature squeezing using practical examples. Sec. 3 provides a detailed description of the experimental setup. Sec. 4 presents the experimental results on polarization squeezing using intense pulses. Two different beam combinations are considered which lead to the generation of distinct nonclassical two-modes states: 1) A linearly polarized intense amplitude squeezed beam is combined with a vacuum resulting in sub-shot-noise Stokes variances but not in polarization squeezing; and 2) Polarization squeezed light emerging from the combination of two orthogonally polarized intense amplitude squeezed beams. Sec. 5 is devoted to the conclusions.

\section{Theory}

The classical Stokes parameters are a well known description of the polarization state of light \cite{Sto52,BorWol99}. Of interest to this paper are their quantum counterparts. In direct analogy to the classical parameters, we find that the quantum Stokes operators are \cite{JauRoh55,KorLeu01a} and references therein:
\begin{eqnarray}
	\hat{S}_{0} &=& \hat{a}^{\dagger}_{x} \hat{a}_{x} + \hat{a}^{\dagger}_{y} \hat{a}_{y} = \hat{n}_x + \hat{n}_y = \hat{n}, \nonumber \\
	\hat{S}_{1} &=& \hat{a}^{\dagger}_{x} \hat{a}_{x} - \hat{a}^{\dagger}_{y} \hat{a}_{y} = \hat{n}_x - \hat{n}_y, \nonumber \\
	\hat{S}_{2} &=& \hat{a}^{\dagger}_{x} \hat{a}_{y} + \hat{a}^{\dagger}_{y} \hat{a}_{x}, \nonumber \\
	\hat{S}_{3} &=& i(\hat{a}^{\dagger}_{y} \hat{a}_{x} - \hat{a}^{\dagger}_{x} \hat{a}_{y}),
	\label{eqn:stokes_def}
\end{eqnarray}
where $\hat{a}_{x/y}$ and $\hat{a}^{\dagger}_{x/y}$ refer to the photon annihilation and creation operators of two orthogonal polarization modes x and y. $\hat{n}_x$ and $\hat{n}_y$ are the photon number operators of these modes and $\hat{n}$ is the total number operator. The $\hat{S}_{0}$ operator corresponds to the beam intensity whilst $\hat{S}_{1}$, $\hat{S}_{2}$ and $\hat{S}_{3}$ describe the polarization state. The $\hat{S}_0$ operator commutes with the others:
\begin{eqnarray}
  [\hat{S}_0,\hat{S}_j] = 0, \quad j = 1,2,3,
  \label{eqn:stokes_commutator}
\end{eqnarray}
whereas the remaining operators obey the SU(2) Lie algebra, as indicated by the commutator:
\begin{eqnarray}
  [\hat{S}_1,\hat{S}_2] = 2i\hat{S}_3,
  \label{eqn:stokes_commute}
\end{eqnarray}
and the cyclics thereof. These non-zero commutators imply the impossibility of simultaneous exact measurement of the operators. It is worth noting that the commutators are operator valued. The commutation relations here are similar to those for spin $\frac{1}{2}$ particles and this fact has already been experimentally leveraged in the mapping of the quantum polarization state of light onto an atomic spin system \cite{HalSor01}. As a result of Eq.~(\ref{eqn:stokes_commute}) uncertainty relations in the variances of the operators arise:
\begin{equation}
V_{1}V_{2}\geq{}|\langle\hat{S}_3\rangle|^{2}, \quad{}V_{3}V_{1}\geq{}|\langle\hat{S}_2\rangle|^{2}, \quad{}V_{2}V_{3}\geq{}|\langle\hat{S}_1\rangle|^{2}.
\label{eqn:stokes_uncertainty}
\end{equation}
Here $V_j$ refers to the variance $\langle\hat{S}_j^2\rangle - \langle\hat{S}_j\rangle^2$ of the Stokes operator $\hat{S}_j$.

\begin{figure}
	\begin{center}
		\includegraphics[scale=0.9]{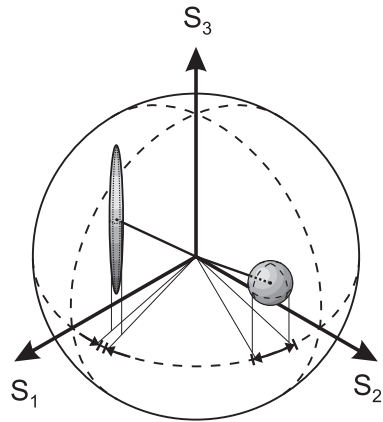}
	\end{center}
	\caption{The Poincar\'e sphere with a coherent state (sphere) and a polarization squeezed beam (cigar).}
	\label{fig:gen_poincare}
\end{figure}

A useful aid in visualizing a polarization state and its uncertainties is the quantum Poincar\'e sphere \cite{KorLeu01a}. It is defined as:
\begin{eqnarray}
  \hat{S}_1^2 + \hat{S}_2^2 + \hat{S}_3^2 = \hat{S}_0^2 + 2\hat{S}_0.
\end{eqnarray}
This differs from the classical definition \cite{BorWol99} in the 2$\hat{S}_0$ term which is a result of the non-commutation of the Stokes operators. On the Poincar\'e sphere uncertainty regions are depicted as volumes about the mean values of the operators, as shown in Fig.~\ref{fig:gen_poincare}. The modes depicted there are: 1) A spherical uncertainty volume of radius $\sqrt{3\langle n \rangle}$ centered on the radial value of $\langle n \rangle$ representing an arbitrarily polarized coherent beam; 2) A cigar shaped uncertainty which is squeezed below the shot noise in $\hat{S}_0$, $\hat{S}_1$ and $\hat{S}_2$ and anti-squeezed in $\hat{S}_3$ depicting an arbitrarily chosen polarization squeezed beam. 

\begin{figure}
	\begin{center}
		\includegraphics[scale=0.67]{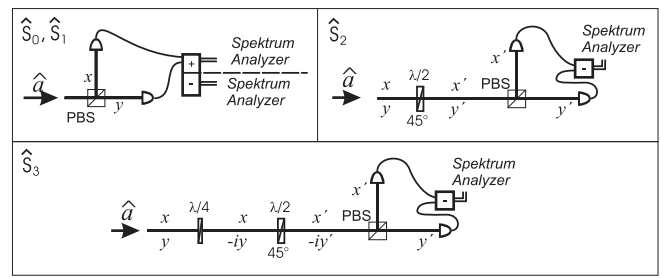}
	\end{center}
	\caption{Measurement schemes for the Stokes operators' variances.}
	\label{fig:stokes_detect}
\end{figure}

A particular advantage of the Stokes operators is that they can be directly measured using only linear optical elements thereby avoiding local oscillator methods \cite{KarMas93,KorLeu01a, AgrCha02}. The required setups are shown in Fig.~\ref{fig:stokes_detect}. Measuring the difference channel in balanced detection gives the $\hat{S}_1$ parameter. Addition of a half wave-plate such that the incoming polarization is rotated by 45$^\circ$ allows for the measurement of the $\hat{S}_2$ operator. Introduction of a quarter wave-plate to the $\hat{S}_2$ setup such that the wave-plate's axes coincide with those of the polarization beam splitter permits measurement of the $\hat{S}_3$ parameter. The $\hat{S}_0$ operator is found in all three setups by simply measuring the sum channel, though the first is taken as the standard. This freedom arises from the fact that $\hat{S}_0$ represents the total intensity (c.f. Eq.~(\ref{eqn:stokes_def})) and as such is constant for all configurations.

The three operator valued commutation relations of the Stokes operators make the definition of polarization squeezed states such as that in Fig.~\ref{fig:gen_poincare} non-trivial. In contrast to the quadrature operators \cite{Lou00}, the coherent polarization state is not simultaneously a minimum uncertainty state (MUS) for all Stokes operators. Considering the Stokes uncertainty relations in Eq.~(\ref{eqn:stokes_uncertainty}), the MUS is represented by an equality sign "=" in place of the "$\geq$":

\begin{equation}
V_{1}V_{2}=|\langle\hat{S}_3\rangle|^{2},\quad{}V_{3}V_{1}=|\langle\hat{S}_2\rangle|^{2},\quad{}V_{2}V_{3}=|\langle\hat{S}_1\rangle|^{2}.
\label{eqn:stokes_MUS}
\end{equation}

The mean values of the third Stokes operator on the right of the equations indicate the state dependence of the MUS of the other two operators. That is, the specific form of the uncertainty relations and the MUS are determined by the particular polarization state under consideration.

As we investigate the definition of polarization squeezing let us first consider the quadrature MUS, the coherent mode:
\begin{eqnarray}
  |\psi\rangle = |\alpha_x\rangle\alpha_y\rangle,
\end{eqnarray}
with the Stokes variances:
\begin{eqnarray}
  V^{coh}_j = \langle \hat{n}_x \rangle + \langle \hat{n}_y \rangle = \langle \hat{n} \rangle, & j = 1,2,3.
\end{eqnarray}
This incorrectly leads to what seems to be a natural definition for polarization squeezing in analogy to that for quadrature squeezing:
\begin{eqnarray}
  V_j < V^{coh}_j = \langle \hat{n} \rangle, & j=1,2,3.
  \label{eqn:coh_polsq_def}
\end{eqnarray}
This suggest that a polarization squeezed state is one in which the variance of a Stokes operator $\hat{S}_j$ falls below the variance of an equally intense coherent mode. A polarization state obeying Eq.~(\ref{eqn:coh_polsq_def}) is always a nonclassical state and can be used to produce quadrature entanglement. However, this definition alone implies nothing more than conventional quadrature or single-mode squeezing observed through the measurement of the Stokes parameter. 

The correct definition of polarization squeezing must consider the conditions given by the uncertainty relations of Eq.~(\ref{eqn:stokes_uncertainty}) and the MUS bounds of Eq.~(\ref{eqn:stokes_MUS}). These show that the coherent state is not a polarization MUS for any pair of Stokes operators. For example, the inequality of Eq.~(\ref{eqn:coh_polsq_def}) can hold simultaneously for a pair of Stokes operators for which the uncertainty relations are bounded from below by zero. They can then both be measured simultaneously and exactly and the usual concept of squeezing is not applicable. Therefore, to define a meaningful nonclassical polarization state the uncertainty relations and the MUS must be taken into account. Thus we define a polarization squeezed state as one in which one of the Stokes variances lies not only below the coherent limit but also below the respective MUS limit (c.f. Eq.~(\ref{eqn:stokes_MUS})) \cite{KorLou03}:
\begin{eqnarray}
  V_j < |\langle \hat{S}_l \rangle | < V_k , \quad j \neq k \neq l.
  \label{eqn:polsq_def}
\end{eqnarray}

In this paper the term polarization squeezing is always used in the strict sense of the above definition, Eq.~(\ref{eqn:polsq_def}). In contrast to Eq.~(\ref{eqn:coh_polsq_def}) this definition is closely related to two-mode squeezing. Note that some papers have used the term in the looser sense of Eq.~(\ref{eqn:coh_polsq_def}) \cite{KorLeu01a,BowSch02}. The implications of the definition in Eq.~(\ref{eqn:polsq_def})and its differences compared to Eq.~(\ref{eqn:coh_polsq_def}) will be further investigated by way of three examples.

Before beginning discussion of these examples, the amplitude ($\hat{X}^+$) and phase ($\hat{X}^-$) quadrature operators should be introduced. These are useful tools to describe the quantum state of light and are particularly advantageous here as our polarization squeezing is generated from amplitude quadrature squeezing. Thus the operators are \cite{Lou00}:
\begin{equation}
\hat{X}_j^+=\hat{a}_j^\dagger+\hat{a}_j,\quad{}\hat{X}_j^-=i(\hat{a}_j^\dagger-\hat{a}_j),\quad{}j=x, y,
\label{eqn:quadrature_def}
\end{equation}
using the notation of Eq.~(\ref{eqn:stokes_def}). These operators obey the uncertainty relation:
\begin{eqnarray}
  V(\hat{X}_j^+) V(\hat{X}_j^-) \geq 1,
  \label{eqn:quadrature_unc}
\end{eqnarray}
which implies the impossibility of simultaneous exact measurement of the operators. For easier use of the quadrature operators we can describe the photon operators by:
\begin{eqnarray}
  \hat{a}_j = \alpha_j + \delta\hat{a}_j,
  \label{eqn:linearisation}
\end{eqnarray}
where $\alpha$ is the real, classical amplitude of the mode $j$ and $\delta\hat{a}_j$ is the corresponding quantum noise operator with a mean value of zero. Upon substitution into Eq.~(\ref{eqn:quadrature_def}) we find the quadrature noise operators are:
\begin{eqnarray}
  \delta\hat{X}^+_j = \delta\hat{a}_j^\dagger + \delta\hat{a}_j, \quad \delta\hat{X}^-_j = i(\delta\hat{a}_j^\dagger - \delta\hat{a}_j).
\end{eqnarray}
For an amplitude squeezed beam the variance, $V(\delta\hat{X}_j^+)$ is decreased relative to the coherent state whilst $V(\delta\hat{X}_j^-)$ is increased, or anti-squeezed. Using this basis we will now look at three forms of polarization squeezing arising from different combinations of amplitude squeezed and coherent beams. Their investigation will show why for viable application of polarization squeezing of intense light, two amplitude squeezed beams are necessary.

\subsubsection{Example 1}

The first case we will examine is that of a single, linearly x-polarized, amplitude squeezed beam combined where the y-polarization mode is represented by a coherent vacuum. These states are described in the following box:
\begin{center}
\begin{tabular}[t]{|c|c|} \hline \textbf{x--polarization mode} & \textbf{y--polarization mode} \\ 
\hline amplitude squeezed & coherent vacuum \\ 
\hline ${\hat{a}}_x= \alpha + \delta{\hat{a}}_x$ & ${\hat{a}}_y= \delta {\hat{a}}_y$ \\ 
\hline $V(\delta X^{+}_{x})<1$, $V(\delta X^{-}_{x})>1$ & $V(\delta X^{+}_{y})=V(\delta X^{-}_{y})=1$ \\
\hline
\end{tabular} \end{center}
Here the  beams have been described as in Eq.~(\ref{eqn:linearisation}). The resultant variances for the amplitude and phase quadratures are also shown. The following mean Stokes values are found for this beam:
\begin{equation}
\begin{array}{rl}
\langle\hat{S_{0}}\rangle={\alpha}^{2}_{x}, & \quad\langle\hat{S_{1}}\rangle={\alpha}^2_{x}, \\
\langle\hat{S_{2}}\rangle=0, & \quad\langle\hat{S_{3}}\rangle=0.
\end{array}
\end{equation}
In the spirit of squeezing relative the coherent beam (see Eq.~(\ref{eqn:coh_polsq_def})), we normalize the variance with respect to the coherent state: $\tilde{V}_{j} = V_{j} / V^{coh}_j = V_{j} / \langle \hat{n} \rangle$. Thus a normalized variance of 1 corresponds to a coherent state and a value of less than one indicates squeezing. Correspondingly, the coherent amplitude is normalized to 1. Substituting the above mean Stokes values into the Stokes uncertainty relations, Eq.~(\ref{eqn:stokes_uncertainty}), we find:

\begin{eqnarray}
\begin{array}{ccc} {{\tilde{V}}_{1}{\tilde{V}}_{2}\geq{}|\langle\hat{S}_3\rangle|^2=0,} & &
{\tilde{V}}_{2}{\tilde{V}}_{3}\geq{}|\langle\hat{S}_1\rangle|^2=1, \\ \multicolumn{3}{c}{{{\tilde{V}}_{3} {\tilde{V}}_{1} \geq  |\langle \hat{S}_2 \rangle|^2  = 0. }} \end{array}
\label{eqn:ex1_uncertainties}
\end{eqnarray}

First, we see that the operator pairs $\hat{S_{1}}$ \& $\hat{S_{2}}$ and $\hat{S_{1}}$ \& $\hat{S_{3}}$ commute, as the uncertainty relations are bounded by zero. Thus values for the operator pairs can be obtained simultaneously and exactly. Second, the variables $\hat{S_{2}}$ and $\hat{S_{3}}$ are conjugate as their uncertainty relation is bounded from below by a non-zero limit. That is their values can not be simultaneously measured to an accuracy better than the state dependent limit. This system of inequalities is similar to the usual Heisenberg uncertainty relation, for example of position and momentum, inasmuch as only one uncertainty relation has a non-zero value. The state's MUS is found to be when the "$\geq$" are replaced by "=". Squeezing of one of the conjugate variables below this bound results in Heisenberg-like squeezing which requires the conjugate variable to increase correspondingly, i.e. ${\tilde{V}}_2 < 1 < {\tilde{V}}_3$.

The products of these variances can also be calculated from the quadrature variances given in the above box. We find the following Stokes variances:
\begin{eqnarray}
 {\tilde{V}}_0 = V(\delta X^{+}_{x}) < 1, & & \quad {\tilde{V}}_1 = V(\delta X^{+}_{x}) < 1, \nonumber \\
 {\tilde{V}}_2 = 1, & & \quad {\tilde{V}}_3 = 1.
\end{eqnarray}
These give rise to the following variance products:
\begin{eqnarray}
  {\tilde{V}}_{1} {\tilde{V}}_{2} \leq 1, \quad {\tilde{V}}_{2} {\tilde{V}}_{3} = 1, \quad {\tilde{V}}_{3} {\tilde{V}}_{1} \leq 1.
  \label{eqn:ex1_varproducts}
\end{eqnarray}
These equations show the behavior of the system relative to the coherent state. In particular we see that $\hat{S}_{0}$ and $\hat{S}_{1}$ are squeezed in the sense of Eq.~(\ref{eqn:coh_polsq_def}), i.e. quadrature squeezing. Since the variances of both conjugate variables, $\hat{S}_{2}$ and $\hat{S}_{3}$, are equal to 1, this state does not exhibit polarization squeezing. Additionally, we see that this state can only be a minimum uncertainty state in the limit of infinite squeezing ($V(\delta X^{+}_{x}) \rightarrow 0$). This demonstrates that merely adding vaccum modes to a quadrature squeezed mode does not lead to truly multi-mode effects such as polarization squeezing.

\subsubsection{Example 2}

In this example a linearly x-polarized, amplitude squeezed beam is overlapped with a linearly y-polarized, equally intense, coherent beam.
\begin{center}
\begin{tabular}[t]{|c|c|} 
\hline \textbf{x--polarization mode} & \textbf{y--polarization mode} \\ 
\hline amplitude squeezed & coherent state\\ 
\hline ${\hat{a}}_x= \alpha + \delta {\hat{a}}_x$ & ${\hat{a}}_y=\alpha + \delta {\hat{a}}_y$ \\ 
\hline $V(\delta X^{+}_{x})<1$, $V(\delta X^{-}_{x})>1$ & $V(\delta X^{+}_{y})=V(\delta X^{-}_{y})=1$ \\
\hline
\end{tabular} \end{center}
This gives rise to the mean Stokes values seen below:
\begin{eqnarray}
  \langle \hat{S_{0}} \rangle = 2{\alpha}^{2}, & \quad \langle \hat{S_{1}} \rangle = 0, \nonumber \\
  \langle \hat{S_{2}} \rangle = 2 {\alpha}^2,  & \quad \langle \hat{S_{3}} \rangle = 0,
\end{eqnarray}
which lead to the following uncertainty relations:
\begin{eqnarray}
  {\tilde{V}}_{1} {\tilde{V}}_{2} \geq 0, \quad {\tilde{V}}_{2}{\tilde{V}}_{3} \geq 0, \quad {\tilde{V}}_{3} {\tilde{V}}_{1} \geq 1.
  \label{eqn:ex2_uncertainty}
\end{eqnarray}

Again we finding commuting operator pairs: $\hat{S_{1}}$ \& $\hat{S_{2}}$ and $\hat{S_{2}}$ \& $\hat{S_{3}}$. In this example as a result of $\hat{S}_2$ being non-zero rather than $\hat{S}_1$ the conjugate variables are $\hat{S_{1}}$ and $\hat{S_{3}}$. Now considering the squeezing variances in the above box, the normalized variances are:
\begin{eqnarray}
 {\tilde{V}}_0 = \frac{V(\delta X^{+}_{x}) +1}{2} < 1, \quad {\tilde{V}}_1 = \frac{V(\delta X^{+}_{x}) +1}{2} < 1, \nonumber \\
 {\tilde{V}}_2 = \frac{V(\delta X^{+}_{x}) +1}{2} < 1, \quad {\tilde{V}}_3 = \frac{V(\delta X^{-}_{x}) +1}{2} > 1.
 \label{eqn:ex3var}
\end{eqnarray}
Considering the uncertainty relation for the quadrature operators in Eq.~(\ref{eqn:quadrature_unc}) in conjunction with the variance above we find the following variance products:
\begin{eqnarray}
  {\tilde{V}}_{1} {\tilde{V}}_{2} < 1, \quad {\tilde{V}}_{2} {\tilde{V}}_{3} \geq 1, \quad {\tilde{V}}_{3} {\tilde{V}}_{1} \geq 1.
  \label{eqn:ex2_varproduct}
\end{eqnarray}
In contrast to example 1, not only are both $\hat{S_{0}}$ and $\hat{S_{1}}$ squeezed relative the corresponding coherent state, but $\hat{S_{2}}$ is also. Again two of the parameters are conjugate -- $\hat{S_{1}}$ and $\hat{S_{3}}$. 

They are indeed found to obey a Heisenberg-like uncertainty relation and thus polarization squeezing as in Eq.~(\ref{eqn:polsq_def})is possible, i.e. ${\tilde{V}}_1 < 1 < {\tilde{V}}_3$. It must be noted that, despite seeing squeezing in three of the four Stokes parameters, this squeezing is small with respect to the initial amplitude squeezing value. This is because of the mixing with the coherent state, the effect of which is in Eq.~(\ref{eqn:ex3var}). For example, 3~dB squeezing in $V(\delta X_x^+)$ gives only 1.25~dB squeezing in $\hat{S}_{0}$, $\hat{S}_{1}$ and $\hat{S}_{2}$. Also of interest is that this beam would exhibit two-mode squeezing if incident on a polarising beam splitter at $45^\circ$ to x and y basis.

\subsubsection{Example 3}

In the final example we investigate a case similar to example 2, but here both beams are amplitude squeezed with equal squeezing and intensity magnitudes.
\begin{center}
\begin{tabular}[t]{|c|c|} 
\hline \textbf{x--polarization mode} & \textbf{y--polarization mode} \\ 
\hline amplitude squeezed & amplitude squeezed\\ 
\hline ${\hat{a}_x}= \alpha + \delta{\hat{a}}_x$ & ${\hat{a}}_y=\alpha + \delta {\hat{a}}_y$ \\
\hline $V(\delta X^{+}_{x})<1$, $V(\delta X^{-}_{x})>1$ & $V(\delta X^{+}_{y}) <1, V(\delta X^{-}_{y})>1$ \\
\hline
\end{tabular} 
\end{center}
The resultant mean Stokes values are:
\begin{eqnarray}
  \langle \hat{S_{0}} \rangle = 2{\alpha}^{2}, \quad \langle \hat{S}_{1} \rangle = 0, \nonumber \\
  \langle \hat{S_{2}} \rangle = 2 {\alpha}^2, \quad \langle \hat{S_{3}} \rangle = 0,
\end{eqnarray}
which give rise to identical uncertainty relations as in example 2 (see Eq.~(\ref{eqn:ex2_uncertainty})). The normalized Stokes variances derived from the two amplitude squeezed beams are:
\begin{eqnarray}
 {\tilde{V}}_0 = V(\delta X^{+}) < 1, \quad {\tilde{V}}_1 = V(\delta X^{+}) <1, \nonumber \\
 {\tilde{V}}_2 = V(\delta X^{+}) < 1, \quad {\tilde{V}}_3 = V(\delta X^{-}) > 1,
\end{eqnarray}
where the variance products are similar to those in example 2 (see Eq.~(\ref{eqn:ex2_varproduct})). We have made use of the fact that the beams exhibit identical squeezing: $V(\delta X^{+}_{x})=V(\delta X^{+}_{y})=V(\delta X^{+})$ and $V(\delta X^{-}_{x})=V(\delta X^{-}_{y})=V(\delta X^{-})$. These results parallel those in example 2 and polarization squeezing as in Eq.~(\ref{eqn:polsq_def}) is again observed in $\hat{S}_{1}$ with $\hat{S}_3$ as its anti-squeezed conjugate. The important difference here with respect to example 2 is that $\hat{S}_0$, $\hat{S}_1$ and $\hat{S}_2$ are squeezed to the same extent as the individual amplitude squeezed beams. Thus this case holds the most promise for quantum information and so is the primary focus of this paper.

\begin{figure}[h]
	\includegraphics[scale=0.65]{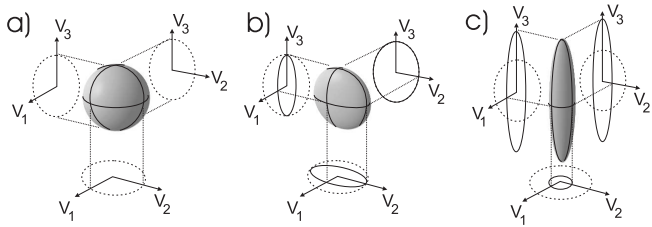}
	\caption{Uncertainty shapes for a) a coherent state, b) a single amplitude squeezed beam, and c) a state composed of two orthogonally polarized, phase-matched, amplitude squeezed beams.}
	\label{fig:3dpoincare}
\end{figure}

Fig.~\ref{fig:3dpoincare} illustrates the variances of the different polarization squeezed states that have been discussed. The coherent state is represented by the spherical uncertainty volume in Fig.~\ref{fig:3dpoincare}a. The state Fig.~\ref{fig:3dpoincare}b corresponds to the example 1, an amplitude squeezed beam. As such, only squeezing below the coherent limit is seen and no anti-squeezing is present, indicating no polarization squeezing. Examples 2 and 3 are described by Fig.~\ref{fig:3dpoincare}c, where we see squeezing in $\hat{S}_1$ and $\hat{S}_2$ relative the coherent bound. $\hat{S}_1$ is polarization squeezed as it is a conjugate variable squeezed below its uncertainty limit, and $\hat{S}_3$ is correspondingly anti-squeezed.

To maximize the available polarization squeezing in our experiments, we use the superposition of two amplitude squeezed beams for the generation of polarization squeezed light, as in example 3. The above calculations assume a perfect phase alignment between the x- and y-polarization modes. This raises the question as to what effect an imperfect phase match between the pulses has on the squeezing. If we assume an angle $\phi$ between the orthogonally polarized modes, they are described by:
\begin{eqnarray} \left(
\begin{array}{clrr}
	\hat{a}_{x} \\ \hat{a}_{y}
\end{array} \right) = \left(
\begin{array}{clrr}
  \alpha_{x} + \delta\hat{a}_{x} \\
  e^{i\phi}(\alpha_{y} + \delta\hat{a}_{y})
\end{array} \right). \end{eqnarray}
Using this to calculate the Stokes operators we find:
\begin{eqnarray}
   \hat{S}_{0} &=& \alpha^{2}_{x} + \alpha_{x}\delta\hat{X}^{+}_{x} + \alpha^{2}_{y} + \alpha_{y}\delta\hat{X}^{+}_{y}, \nonumber \\
   \hat{S}_{1} &=& \alpha^{2}_{x} + \alpha_{x}\delta\hat{X}^{+}_{x} - (\alpha^{2}_{y} + \alpha_{y}\delta\hat{X}^{+}_{y}), \nonumber \\
   \hat{S}_{2} &=& 2\alpha_{x}\alpha_{y}\cos\phi + \alpha_{y}(\delta\hat{X}^{+}_{x}\cos\phi + \delta\hat{X}^{-}_{x}\sin\phi) +
   \nonumber \\ & & \alpha_{x}(\delta\hat{X}^{+}_{y}\cos\phi - \delta\hat{X}^{-}_{y}\sin\phi), \nonumber \\
   \hat{S}_{3} &=& 2\alpha_{x}\alpha_{y}\sin\phi + \alpha_{y}(\delta\hat{X}^{+}_{x}\sin\phi + \delta\hat{X}^{-}_{x}\cos\phi) + 
   \nonumber \\ & & \alpha_{x}(\delta\hat{X}^{+}_{y}\sin\phi - \delta\hat{X}^{-}_{y}\cos\phi).
   \label{eqn:phase_effect}
\end{eqnarray}
These give the following mean values:
\begin{eqnarray}
  \langle\hat{S}_{0}\rangle &=& \alpha^{2}_{x} + \alpha^{2}_{y}, \nonumber \\
  \langle\hat{S}_{1}\rangle &=& \alpha^{2}_{x} - \alpha^{2}_{y}, \nonumber \\
  \langle\hat{S}_{2}\rangle &=& 2\alpha_{x}\alpha_{y}\cos\phi, \nonumber \\
  \langle\hat{S}_{3}\rangle &=& 2\alpha_{x}\alpha_{y}\sin\phi.
\end{eqnarray}
We see that $\hat{S}_{0}$ and $\hat{S}_{1}$ should be phase insensitive in both mean and noise. However $\hat{S}_{2}$ and $\hat{S}_{3}$ are phase sensitive, making good phase stability a requirement.

\section{Experiment}

\begin{figure}[h]
	\includegraphics[scale=0.67]{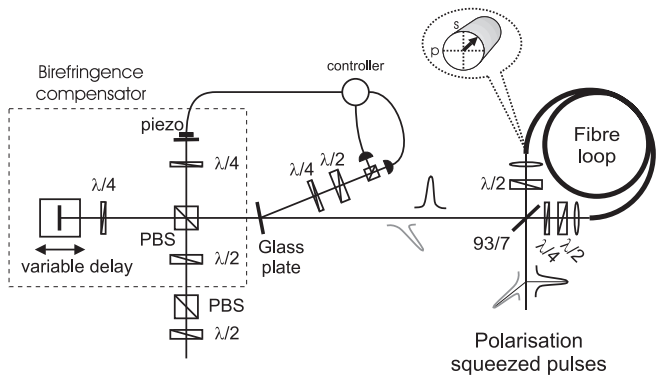}
	\caption{Schematic of the experimental setup.}
	\label{fig:Aufbau}
\end{figure}

In this work we report on an experiment to produce quadrature and polarization squeezed light beams by spatially overlapping an intense, amplitude squeezed pulse with either a coherent vacuum or another amplitude squeezed pulse. Polarization squeezing was recently implemented using quadrature squeezed cw-light from an optical parametric oscillator \cite{BowSch02}. In contrast, our work takes advantage of an efficient, pulsed squeezing source afforded by the Kerr nonlinearity of a glass fiber \cite{RosShe91,BerHau91,SchFic98b,KorLeu00}. The setup used is depicted in Fig.~\ref{fig:Aufbau}.

The experiment uses a passively mode-locked Cr$^{4+}$:YAG laser operated at 1495 nm and a repetition frequency of 163~MHz \cite{SpaBoh97}. The maximum average power is 95~mW and the pulses have a bandwidth limited hyperbolic-secant shape with a FWHM of 130-150~fs. These pulses are squeezed utilizing the Kerr effect ($\chi^{(3)}$ nonlinearity) in an asymmetric fiber-Sagnac interferometer \cite{SchFic98b,KorLeu00}. The Sagnac loop consists of a 93:7 polarization independent beam splitter and 14.2~m of polarization maintaining fiber with negligible polarization cross-talk (FS-PM-7811 from 3M), chosen for its small mode diameter and thus high nonlinearity which allows good squeezing at relatively low pulse energy. The beam splitter divides a pulse into two parts of unequal energy which then counter propagate through the fiber. The strong pulse experiences the fiber nonlinearity much more intensely than the weak pulse, giving rise to an intensity dependent, nonlinear phase shift. In the single mode picture described in \cite{KitYam86} this effect squeezes the circular shaped phase space uncertainty into a crescent. However, reduced amplitude fluctuations can not be observed directly as the deformed uncertainty area is not oriented along the amplitude quadrature direction. Upon interference with the counter propagating weak pulse on the beam splitter after exiting the fiber the ellipse is reoriented in phase space. For certain input energies this realignment yields detectable amplitude squeezing \cite{SchFic98b}.

Similar to previous experiments by Silberhorn et al. \cite{KorLeu00,SilLam01}, polarization maintaining fiber was used in the Sagnac interferometer to simultaneously generate two independent, orthogonally polarized, amplitude squeezed pulses. Equal squeezing of these pulses was guaranteed by adjusting the quarter wave-plates in the birefringence compensator, seen in Fig.~\ref{fig:Aufbau}, which fine tuned the intensity of the two polarization modes going into the fiber. Instead of interfering these two squeezed pulses on a beam splitter after rotating one of the polarizations as in \cite{SilLam01}, we have brought the orthogonally polarized, squeezed pulses to spatial and temporal overlap, thereby generating polarization squeezing. Due to the birefringence of the fiber (beat length 1.75~mm at 1495~nm), a path difference between the polarizations is necessary for successful overlap. The birefringence compensator accomplishes this by splitting x- and y-polarizations in a Michelson-interferometer-like setup, introducing a path difference of 1.2~cm. Coarse adjustment of the delay is realized using a micrometer table; fine adjustment of the relative phase is achieved by an actively controlled piezo system. Placing the birefringence compensator before the Sagnac interferometer has the advantage of minimizing losses after the interferometer. This is important because squeezing is loss sensitive.

The piezo's control loop is based on the back-reflection from the Sagnac interferometer. This reflection consists of two components: 1) Light reflected from the optical elements without passing through the fiber, and 2) Light which has passed through the fiber and is returned from the interferometer's beam splitter, which has the same polarization state as the light entering the measurement setup. These two signals are of approximately equal intensity because the fiber ends are not anti-reflection coated. The reflected light of 1) is linearly polarized in contrast to the slightly elliptical light which exits the fiber, 2). A $\hat{S}_3$ (ellipticity) measurement is carried out on the back-reflection, and so the reflected light of 1) plays no role. The value of this $\hat{S}_3$ measurement is minimized by the control loop. The loop operates up to several tens of Hertz, constrained by the piezo.

A pair of balanced detectors with windowless InGaAs photo diodes from Epitaxx (ETX-500) were used to measure the Stokes parameters (cf. Fig.~\ref{fig:stokes_detect}). Saturation of the alternating current (AC) amplifier was avoided by suppression of the 163~MHz laser repetition rate and harmonics using a Chebyshev low pass filter. Measurements of the variances of the Stokes parameters, the sum and difference of the AC signals of the two detectors, were made on a pair of spectrum analyzers. These were operated at 17.5~MHz where the extinction ratio of the detectors was 30~dB. Further, a spectrum analyzer resolution bandwidth of 300~kHz, a video bandwidth of 300~Hz and a sweep time of 10~s were used. 

\begin{figure}
	\includegraphics{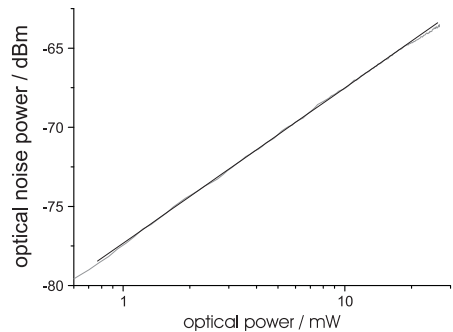}
		\caption{Characteristic of detector.}
	\label{fig:det_nonlin}
\end{figure}

The shot noise for each measurement was determined by the direct current (DC) signals from the detectors which were simultaneously recorded by a digitizing oscilloscope. These DC values were summed after taking into account the slight DC gain difference (1:1.005). From this value the corresponding AC noise was derived. This calculation was based on a linear fit to an AC noise power against DC signal plot for a coherent light beam falling an individual detector. The result is however not absolutely linear, as seen in Fig.~\ref{fig:det_nonlin}. For low input optical powers deviation due to the detector's dark noise and imperfect components was seen. For large input powers the nonlinear behavior is caused by saturation of the AC amplifier. Thus a fit was made to the middle data region. For these reasons, this calibration is the largest source of error in the experiment, conservatively estimated to be 0.2~dB.

\section{Results}
\begin{figure*}
	\includegraphics{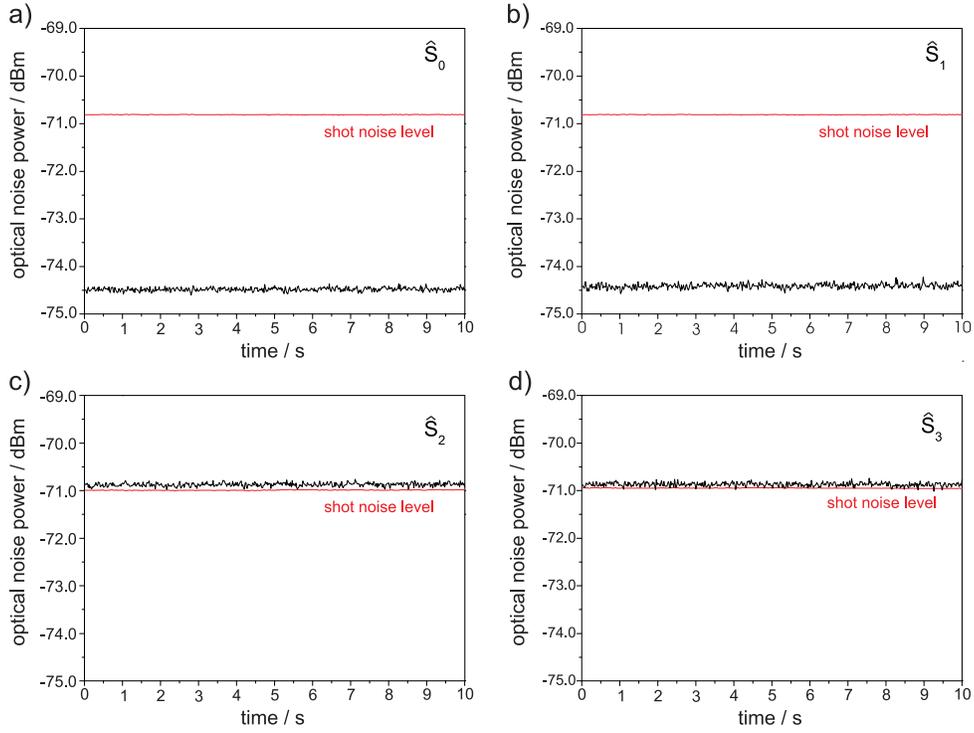}
	\caption{The variances of the Stokes parameters: a) $\hat{S}_{0}$, b) $\hat{S}_{1}$, c) $\hat{S}_{2}$ and d) $\hat{S}_{3}$ for a bright, amplitude squeezed pulse (i.e. example 1) measured over 10 s; the subtracted electronic noise was at -86.2~dB.}
	\label{fig:Stokes_amp_sq_lin2}
\end{figure*}

We used the above described setup to investigate the cases outlined in examples 1 and 3. The results shown in Fig.~\ref{fig:Stokes_amp_sq_lin2} are for example 1: a x-polarized, amplitude squeezed pulse combined with a coherent vacuum on the polarization beam splitter in the measurement setup. As expected, both $\hat{S}_{0}$ and $\hat{S}_{1}$ are squeezed relative the corresponding coherent light limit: by -3.7 and -3.6~dB respectively. However no polarization squeezing as in Eq.~(\ref{eqn:polsq_def})is seen. The remaining parameters, $\hat{S}_{2}$ and $\hat{S}_{3}$, are +0.1~dB above the coherent level. This extra noise is primarily due to the detector and their electronics. The error on these values is $\pm$0.3~dB, resulting from the measurement accuracy and the shot noise calculation. Similar results were seen for 45$^\circ$ and circular polarized beams, whereby the squeezed parameters were $\hat{S}_{0}$ \& $\hat{S}_{2}$ and $\hat{S}_{0}$ \& $\hat{S}_{3}$ respectively. The shot noise levels differ slightly between the measurement runs because of laser power fluctuations as well as detuning of the setup, in particular the fiber coupling.

\begin{figure*}
	\includegraphics{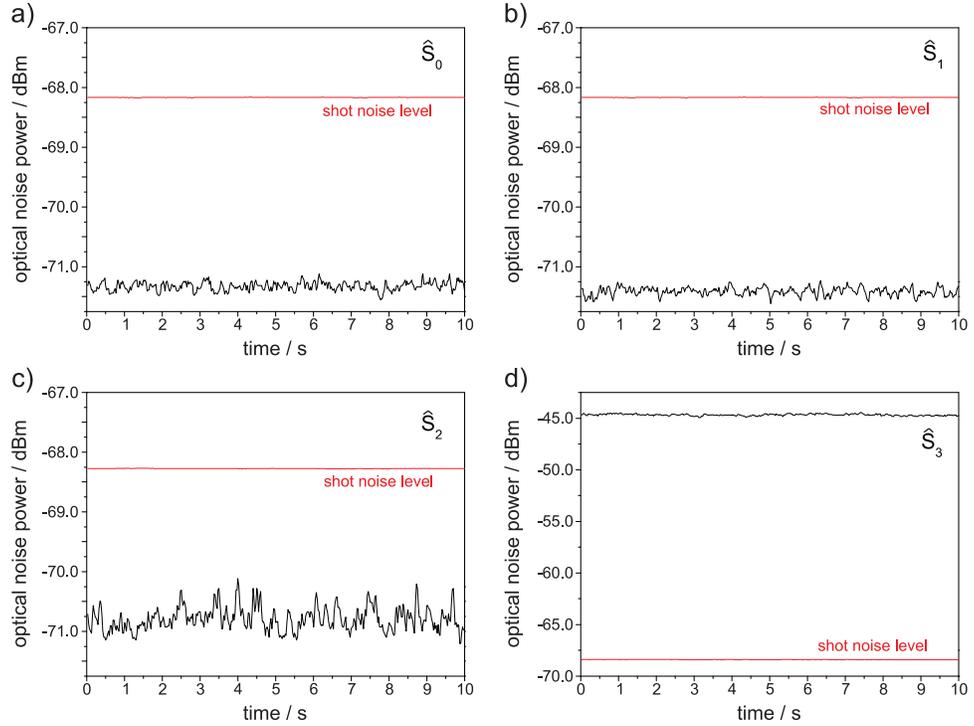}
	\caption{The variances of the Stokes parameters: a) $\hat{S}_{0}$, b) $\hat{S}_{1}$ (polarization squeezed), c) $\hat{S}_{2}$ and d) $\hat{S}_{3}$ for two phase locked, orthogonally polarized, equally bright, similarly amplitude squeezed pulses (i.e. example 3) measured over 10 s; the subtracted electronic noise was at -86.2~dB.}
	\label{fig:pol_sq_zeit2}
\end{figure*}

The case outlined in example 3 -- the overlap of two equivalent, amplitude squeezed pulses -- was also experimentally investigated. The results thereof are dispalyed in Fig.~\ref{fig:pol_sq_zeit2}. In these data series the relative phase ofthe pulses was locked to $\phi$=0 giving rise to a linearly polarized beam at 45$^\circ$. Here three of the four Stokes paarameters are squeezed: $\hat{S}_0$ and $\hat{S}_1$ by -3.4~dB and $\hat{S}_2$ by -2.8~dB. $\hat{S}_1$ is polarization squeezed, as its noise value has been brought under the corresponding minimum bound. The conjugate anti-squeezed parameter is $\hat{S}_3$ with a value of +23.5~dB. It is so large as it also bears the additional phase noise inherent to squeezing in fibers. It should be remembered that the squeezing in $\hat{S}_0$ and $\hat{S}_2$ is only quadrature squeezing.

The general increase in noise for two pulses compared with the single pulse case has three primary roots. The first is increased sensitivity to slight misalignments and general imperfections in the many wave-plates. The misalignment of the wave-plates causes a slight mixing of the difference signal parameters $\hat{S}_{1}$, $\hat{S}_{2}$ and $\hat{S}_{3}$ with each other. Mixing even a small amount of the anti-squeezed parameter, $\hat{S}_{3}$, into the other parameters causes significant squeezing degradation. The sum signal, $\hat{S}_{0}$, is unaffected by this phenomenon. Secondly, fluctuations in the laser power cause significant noise in all parameters. This stems from the fact that squeezing is strongly dependent on pulse power. Further complicating the matter, the four beams coupled into the fiber generally could not be coupled in with identical efficiency giving rise to small squeezing differences between the polarizations. The third error source, the phase noise, affects $\hat{S}_2$ and $\hat{S}_3$, as seen from Eq.~(\ref{eqn:phase_effect}). Whilst $\bar{\phi}$=0, there is significant noise on the beam from thermal and acoustic sources which the phase control can not presently cancel. This error source most greatly affected $\hat{S}_{2}$, which is therefore significantly noisier than the others. The phase noise was in general amplified by the misaligned wave-plates.

Further error sources in the experiment stem from the detectors. Saturation at high intensities was problematic, particularly in the $\hat{S}_{2}$ measurement where the entire beam fell upon one detector. Of the two detectors used one had a noticeably higher saturation level and care was taken to illuminate this detector in the $\hat{S}_{2}$ measurement. The AC gain difference between the detectors, approximately 1:1.06, is not negligible. In a numerical simulation, it was seen that this together with small wave-plate misalignment and the phase noise between the pulses produces significant degradation of the detected squeezing. This effect is strongest for wave-plates standing at 22.5$^\circ$, i.e. for $\hat{S}_{2}$ measurements.

\section{Conclusion}

Quantum information and communication are important fields for the future. Due to the ease of detection of the Stokes operators, polarization squeezing will play an important role in the development of these fields, not least as a basis for polarization entanglement. This paper investigated polarization squeezing of intense pulses using amplitude squeezed pulses generated by the Kerr nonlinearity of a glass fiber. It was seen in three calculated examples that significant squeezing in the Stokes parameters requires stable spatial overlap of two amplitude squeezed beams. This was supported by discussion of experimental results for a single amplitude squeezed beam as well as two overlapped amplitude squeezed beams. In the former only conventional single-mode quadrature squeezing was seen through the Stokes operators whereas the latter exhibited polarization squeezing in $\hat{S}_1$ with $\hat{S}_3$ as its conjugate variable.


\section{Acknowledgements}

The authors gratefully acknowledge the support of the Deutsche Forschungsgemeinschaft and the helpful discussion with F. K\"onig. This project is a part of the EU QIPC Project, No. IST-1999-13071 (QUICOV).


\end{document}